\documentclass[12pt]{article}

\usepackage{amsmath}
\usepackage{amssymb}
\usepackage{slashed}
\usepackage{epsfig}
\usepackage{hyperref}
\usepackage{geometry}
\newcommand{\eps}{\epsilon}
\newcommand{\auth}[3]{{{\large #1}$^{#3}$}\footnote{email:~#2}}
\newcommand{\affS}[2]{{$^{#2}\!\!\!\!$~{\it #1}}}
\newcommand{\verttle}[2][~]{\vspace{-1cm}\begin{flushright}{\small #1}\end{flushright}\vspace{0.5cm} {\sf\bfseries #2}}

\title{\verttle{Production of a KK-graviton and a Vector Boson in ADD Model via Gluon fusion \\}}
\author{
\hspace{-0.8in} \auth{Ambresh Shivaji}{ambresh@iopb.res.in}{1},
  \auth{V. Ravindran}{ravindra@hri.res.in}{2}~and
  \auth{Pankaj Agrawal}{agrawal@iopb.res.in}{1}\\
\hspace{-0.3in}  \affS{ Institute of Physics, Saink School Post, Bhubaneswar 751 005, India}{1} \\
\hspace{-0.3in}  \affS{ Harish-Chandra Research Institute, Chhatnag Road, Jhusi, Allahabad 211 019, India}{2}
}

\begin{document}

\maketitle{}

\begin{center}
 {\bf Abstract}
\end{center}

 {In the models with large 
extra-dimensions, we examine the production of a vector boson ($\gamma$/ $Z$)
in association with the Kaluza-Klein (KK) modes of the graviton via gluon fusion. 
At the leading order, the process takes place through quark-loop box 
and triangle diagrams and it is ultraviolate finite. We report the results 
for the LHC. We also discuss the issues 
of anomaly and decoupling of heavy quarks in the amplitude.}

\section{Introduction}


The problem of large hierarchy between the electroweak and Plank scales can be addressed within 
the models of extra-space dimensions. 
Among various models of extra-dimensions the model of $large$ extra-dimensions, 
known as the ADD model \cite{ADD}, is 
one of the very first successful attempts in this direction.The predictions of 
this model can be probed at present day high energy colliders
 \footnote{Though the ADD model solves the hierarchy problem of the SM, it introduces another 
hierarchy (unexplained within the model) between the scale for the 
size of the extra dimensions ($1/R$) and the fundamental scale of gravity ($M_S$).
 But even then it stands as a good phenomenological model to study.}. 
In the ADD model, the number of space-time dimensions is assumed to be $4+\delta$. 
These $\delta$ extra-space dimensions are considered flat. The standard model (SM) degrees 
of freedom live on a $3+1$ dimensional brane, while gravity is allowed
to propagate in full $4+\delta$ dimensions. To avoid any direct conflict with the
 present day observations, these extra-dimensions are assumed to be compact. 
For simplicity we consider compactification of these extra-dimensions on a 
torus with common scale $R/2\pi$. The fundamental scale of gravity, $M_S$ is related
to the $4-$dimensional Plank scale by

\begin{equation}
 M_P^2 \approx M_S^{\delta+2}\: R^{\delta}.
\end{equation}
 
>From this relation one can argue that if $R$ is large enough the fundamental scale of
 gravity in $4+\delta$-dimensions can be as small as few TeV.
In $4$-dimensions, the $4+\delta$ dimensional graviton appears as an infinite 
tower of KK modes with 
mass spectrum given by, $m_{\vec{n}^2} = \frac{\vec{n}^2}{R^2} $ for
the $n^{th}$ mode. The interaction of spin-$2$ components of these 
KK-modes ($h_{\mu\nu}^{(\vec{n})}$) with the SM is given by an effective Lagrangian

\begin{equation}
\mathcal{L}_{int} \sim \frac{1}{M_P} \sum_{\vec{n}}  h_{\mu\nu}^{(\vec{n})}(x) \: T^{\mu\nu}(x), 
\end{equation} 
where $T^{\mu\nu}$ is the energy-momentum tensor for the SM. Although the cross-section 
for the production of a single KK-graviton
is proportional to $\frac{1}{M_P^2}$, the inclusive cross-section, obtained by 
summing over all KK-modes, is only $\frac{1}{M_S^{\delta+2}} $
suppressed. Thus if $M_S$ is in the TeV range, it can have observable effects at the LHC. Since the coupling of each KK-mode with
the SM particles is very small, the direct production of KK-graviton gives rise to missing energy signal in the detector. More details
on the model and it's phenomenology can be found in \cite{Giudice:1998ck,Mirabelli:1998rt,Han:1998sg}.\\

The direct production of KK-graviton (together with QCD NLO correction) has been considered by many authors in recent past both at the Tevatron and the LHC
\cite{Karg:2009xk,Gao:2009pn,Kumar:2010kv,Kumar:2010ca,Shivaji:2011re}. Due 
to large gluon flux at the LHC the gluon initiated processes can be quite important. In this regard,
 we have investigated
the KK-graviton production in association with a vector boson ($\gamma / Z$) via gluon fusion. 
The paper is organized as follows: In section 2 we introduce our process
and give some details on the structure of the amplitude. Various checks on amplitude and 
the method of computation are described in section 3. 
Numerical results are presented in section 4. We finally conclude in section 5. Some interesting issues 
related to the amplitude calculation are added in the Appendix.

\section{The amplitude}

At the leading order (LO), the process $ gg \rightarrow \gamma/Z+G $ proceeds via quark-loop diagrams. 
The allowed vertices and Feynman rules for the ADD model are derived in Ref. \cite{Han:1998sg}.
Depending upon the coupling of KK-graviton with the standard model particles, 
there are three classes of diagrams.   
The prototype diagram for each class is given in Fig. (1). Other diagrams are obtained just by the 
appropriate permutations of the external legs. There are total of  six box and twelve triangle 
diagrams for each quark flavor. However, due to charge-conjugation property of fermion loop diagrams, 
only half of the diagrams are independent. For $ gg \rightarrow \gamma+G$ case, we find that 
the amplitudes for diagrams related by charge-conjugation are equal and opposite 
to each other. This implies that at the LO

\begin{equation}
 \mathcal{M}(g+g \rightarrow \gamma+G) = 0.
\end{equation}

This is just an implication of the extension of Furry's theorem in the presence of KK-gravitons,
photons, and gluons \cite{mitra}. By introducing charge conjugation transformation of KK-graviton field and 
using charge-conjugation properties of the gluon and photon fields, it can be shown that 
the amplitude does vanish at the LO \footnote{The Graviton field is considered even under charge 
conjugation as it couples with the energy-momentum tensor only. This can be explicitly verified using
the ADD model Lagrangian given in Ref. \cite{Han:1998sg}.}. This result would remain valid to all
orders if only QED and/or QCD radiative corrections are included. In the presence of the weak interaction,
this result would not hold to all orders.\\
  
   For the case of $ gg \rightarrow Z+G $, the amplitude has both vector 
and axial-vector contributions coming from $Z$-boson coupling to quarks.
The vector contribution is similar to $\gamma G$ case in structure. 
Thus, the above argument when applied in $ZG$ case, implies that at the LO 
the process receives contribution only from the axial part of the $Z$-boson coupling. 
For a given quark flavor in the loop the amplitude has the following structure,

\begin{eqnarray}
  \mathcal{M}^{ab}_q(g+g \rightarrow Z+G) = g\; g_s^2\; \kappa\; (\frac{\delta^{ab}}{2})\; c^q_A \; \mathcal{A}(m_q, s, t), \\
\mathcal{A}(m_q,s,t) = \mathcal{A}_{tri}(m_q,s,t)-\mathcal{A}_{box}(m_q,s,t).
\end{eqnarray}
Here $g = \frac{e}{cos\theta_w sin\theta_w }; \; \kappa = \sqrt{16 \pi G_N}; \; c_A^q = - \frac{T_3^q}{2}$.
$s$ and $t$ are Mandelstam variables. Furthermore, $a$ and $b$ are color indices for the two gluons. 
$\mathcal{A}_{tri}$ is the sum of the two types of triangle contributions. 

\begin{figure}[h]
\begin{center}
\includegraphics[width=10cm]{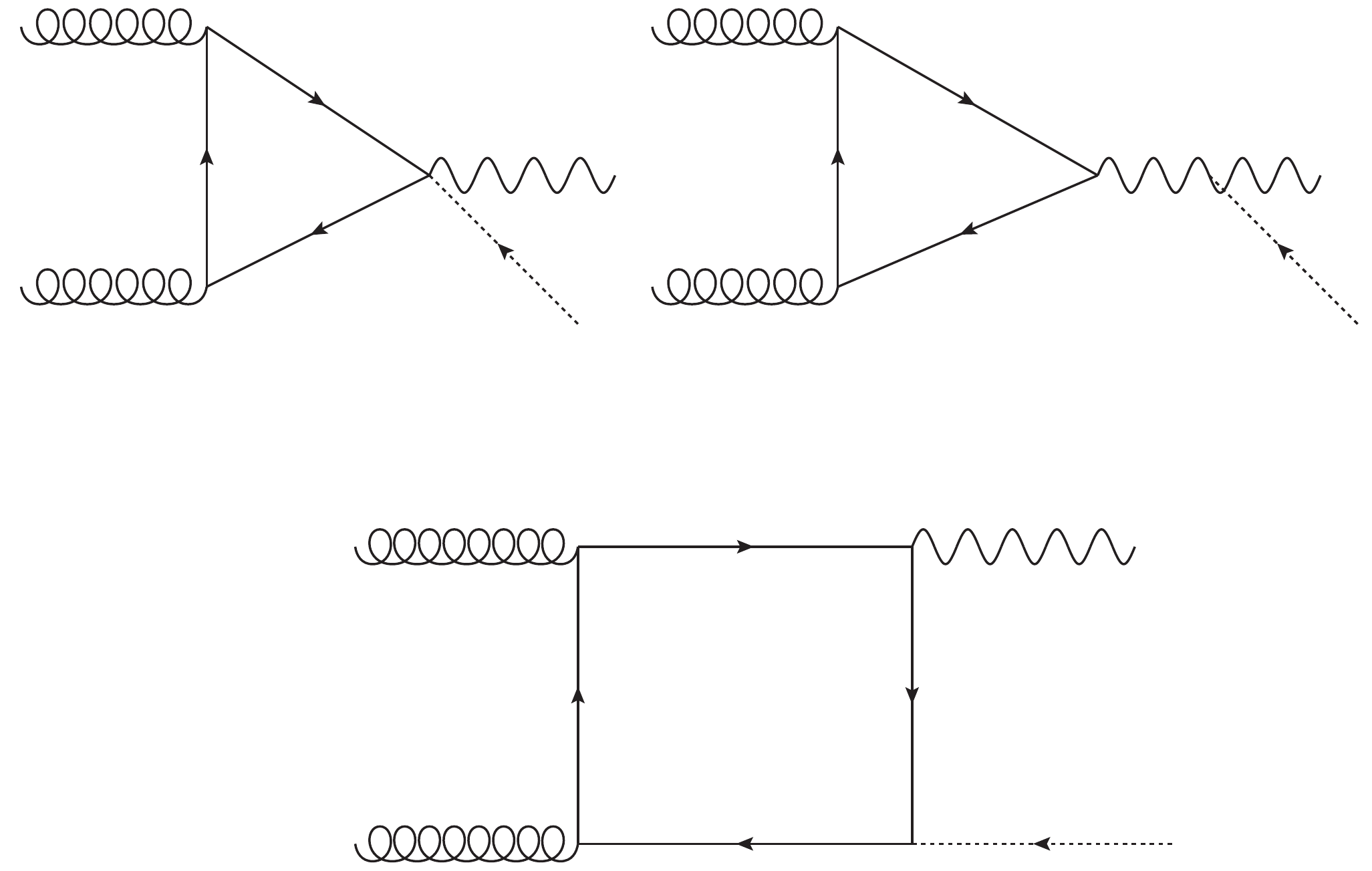}
\caption{Prototype Feynman Diagrams for $ g+g \rightarrow \gamma/Z+G $.}
\end{center}
\end{figure}

This one-loop process is a LO process and is expected to be free of ultraviolet (UV) 
and infrared (IF) singularities for each quark flavour. We also expect gauge invariance
with respect to gluon and KK-graviton current couplings. However, due to the presence of
anomalies, there will be no gauge invariance with respect to axial vector current
coupling for individual quark flavour. But, since the model is free from anomalies, we expect gauge invariance
when we sum over all quark flavours. The confirmation of the cancellation of UV and IR
singularities and gauge invariance with respect to vector and axial-vector current 
couplings are powerful checks on our calculation. We make all these checks as described
in the next section.

Of the six quark flavours, we treat the $u, d, s,$ and $c$ quarks as massless.
 Since the amplitude of the process is proportional to $T_3^q$ value, first 
two generations 
do not contribute. Therefore, the total amplitude, including the contributions
 from all six quarks, is

\begin{eqnarray}
\sum_q  \mathcal{M}_q^{ab}(g+g \rightarrow Z+G) = -\frac{1}{4} g g_s^2 \kappa 
(\frac{\delta^{ab}}{2}) \left[\mathcal{A}(m_t,s,t)-\mathcal{A}(m_b,s,t)\right].
\end{eqnarray}

  We note that the 
cross-section is of $\mathcal{O}(\alpha_s^2)$, and therefore this LO contribution can be included in, 
$\sigma_{NLO}(pp \rightarrow Z+G+X)$ \cite{Kumar:2010kv, Kumar:2010ca}.





\section{Calculation}

  The amplitude for each diagram is written using the Feynman rules given in 
\cite{Han:1998sg}. However, we need not compute all diagrams explicitly. We only need to compute
the prototype diagrams. All other diagrams can be obtained
by suitable permutations of external momenta and polarizations. However, due to the presence of $\gamma_5$ in the
amplitude, one needs to take extreme care in making these permutations in $n$-dimensions. The permutation should not 
be across the $\gamma_5$ vertex. Such permuted diagrams need to be computed explicitly.
Due to the presence of quark-loop, 
each diagram contribution is proportional to a trace over a string of gamma matrices. 
We compute these traces using FORM \cite{Vermaseren:2000nd}. This is the most important
part of the calculation. The presence of $\gamma_5$ in the trace leads to spurious anomalies 
in $4$-dimensions. More details on this is added in appendix.
 We therefore need an appropriate $n$-dimensional treatment of $\gamma_5$. We use Larin's prescription 
for $\gamma_5$ to calculate the trace in $n$-dimensions. According to this prescription

\begin{equation}
 \gamma_\mu \gamma_5 = \frac{i}{6} \eps_{\mu\nu\rho\sigma} \gamma^\nu \gamma^\rho \gamma^\sigma.
\end{equation}

 The fermion loop trace in $n$-dimensions has an interesting feature. It is 
 independent of the fermion mass.
After calculating the trace, we express the amplitude in terms
of appropriate tensor integrals. The box amplitude has rank-four tensor integrals, 
while triangle amplitude has rank-two tensor integrals at the most. At this stage, 
the amplitude in FORM is converted to Fortran routines and is written
in terms of tensor integrals and kinematic invariants. The Fortran routines to
numerically compute these tensor integrals in terms of scalar integrals were 
written in \cite{Agrawal:1998ch} using the methods of Odenbrogh and Vermaseren \cite{vanOldenborgh:1989wn}
\footnote{We have also used an analytical tensor reduction code (in FORM) based on Denner 
and Dittmaier's work \cite{Denner:2002ii}, to make various checks on our amplitude.}. 
The reduction is done in $n=(4-2\eps)$-dimensions. The most general form of amplitude after tensor reduction is

\begin{equation}
 \mathcal{M} = \sum_i \left(d_i D_0^i\right) +  \sum_i \left(c_i C_0^i\right) +  \sum_i \left(b_i B_0^i\right) +  \sum_i \left(a_i A_0^i\right)  + \mathcal{R},  
\end{equation}
where $A_{0}, B_{0}, C_{0}$ and $D_{0}$ are scalar integrals in the notations of 'tHooft and Veltman
\cite{'tHooft:1978xw}. The coefficients in front of scalar integrals (also know as scalar form factors)
contain information on the tensor structure of the amplitude.
$\mathcal{R}$, known as the rational part of the amplitude, is an artifact of UV regularization of 
tensor integrals. For the massless quark case there is 
no $A_0$ term in the above expression \footnote{even for the massive case, the $A_0$ can be replaced by, 
$ A_0(m^2) = m^2(1+B_0(0,m^2,m^2)).$ }. 
We verify that the rational term, $\mathcal{R}$ being an UV effect is independent of the quark mass. \\

All the required scalar integrals for the massless quark case are calculated 
following the method of 'tHooft and Veltman \cite{'tHooft:1978xw}. We need only the 
UV and IR singular pieces of these scalars to verify the cancellation of UV and IR
singularities (which are actually large logs for the case of light quarks).
We regularize UV divergences (present only in $A_0$'s and $B_0$'s) in dimensional regularization,
 while IR divergences are regularized by giving small masses 
to quarks. For the case of top quark and bottom quark, we use FF library to calculate 
the required scalars \cite{vanOldenborgh:1990yc}. 
The above amplitude, after substituting the scalars has following symbolic form

\begin{equation}
 \mathcal{M} = M_{uv}\:\frac{1}{\eps_{uv}} + M_{IR^2} \: ln^2(m^2) +  M_{IR} \: ln(m^2) + M_F .  
\end{equation}

$M_F$ is the finite part of the amplitude and includes the rational term, $\mathcal{R}$. Due to very large and
complicated expression for the amplitude, we compute the amplitude numerically before squaring it. 
This requires computation of polarization vectors for gauge bosons and for the graviton. We have chosen 
helicity basis for them. It also helps in making additional checks on our 
calculation by verifying relations among helicity amplitudes. The polarization tensor for graviton 
is constructed from polarization vectors of massive vector bosons as suggested in \cite{Han:1998sg}.  \\

As discussed in the previous section, the process is expected to be finite. We verify these 
expectations for our amplitude.  Both the massive and massless 
contributions are UV finite. We observe that each triangle diagram is UV finite by itself, 
while Box amplitude is UV finite after adding all the box contributions. Fermion loop 
diagrams are known to be IR finite in both massive and massless fermion cases 
for any kind and any number of external particles attached to the loop \cite{Shivaji:2010aq}.  
In the massless quark case, we check that each diagram is IR finite and therefore 
$M_{IR^2} = M_{IR} = 0$, holds for the full amplitude. In the case of 
triangle class of diagrams involving graviton-gauge boson coupling, we choose $\xi =1 $ (the Feynman gauge)
for gluons case and $\xi = \infty$ (the Unitary gauge) for $Z$-boson case. As expected the calculation does 
not depend on any specific choice of the gauge parameter.
Finally, we check gauge invariance of the amplitude with respect to the two gluons by replacing 
their polarizations with their respective momenta. We observe 
that (only after using $\gamma_5$ prescription
in the trace) both the massless and massive quark contributions are separately gauge invariant 
with respect to the two gluons. 
We also check gauge invariance with respect to the $Z$-boson. Because of the anomaly, 
the two contributions are not separately gauge invariant but the 
total amplitude is gauge invariant up to  the $t$-quark mass as expected due 
to the explicit breaking of the chiral symmetry. All these checks on our amplitude have been 
made both numerically  as well as analytically. The issue of anomaly in the amplitude 
is discussed in appendix.

\section{Numerical Results}

In this section, we present the results of our calculation for the 
process $pp \rightarrow Z + G + X $, at the LHC. The results would
depend on the two parameters of the ADD model - (i) the number of 
extra-space dimensions, $\delta$ and (ii) the fundamental scale of gravity, $M_S$. 
We explore this dependence and other features of the process.
 In Fig. (1), we have 
plotted the cross-section as a function of the center of mass energy.
Here we have chosen $\delta=2$ and $M_S = 2$ TeV. In addition, we
have applied following kinematic cuts:
$$
P_T^Z > 30\; {\rm GeV},\: |\eta^Z| < 2.5 ,\:\sqrt{\hat{s}} < M_S.
$$

The cut on partonic CM energy or equivalently on $ZG$ invariant mass is known as 
{\it truncated scheme} in literature. This cut is related to the fact that theoretical
 predictions within the ADD model, which is an effective theory, 
 can be valid only below the fundamental scale, $M_S$.
It will be interesting to probe the sensitivity of our predictions on this kind of constraint. 
We will further comment on the issue at the end of this section.
We have also chosen factorization and renormalization scales as
$\mu_f = \mu_R = E_T^Z \; (=\sqrt{M_Z^2 + (P_T^Z)^2})$. We have used NLO
parton distribution functions  CTEQ6M. We note that at typical LHC
energy, the cross-section would be of the order of few fb. These cross-sections
are much smaller than expected. The smallness of the cross-section is due
to two-orders of magnitude cancellation in the amplitude between the box-type
and triangle-type diagrams. This destructive interference occurs due to the 
relative minus sign between the box-type and triangle-type
diagrams contributions. Still one may expect few hundred of such events
after a few years of LHC operation at 14 TeV CM energy.\\

 We now illustrate the various aspects of this process at the center of
mass energy of 14 TeV. The parton distribution functions,
factorization/renormalization scales, and the kinematic cuts are as above.


 

\begin{figure}[!h]
\begin{center}
\includegraphics[width=8cm]{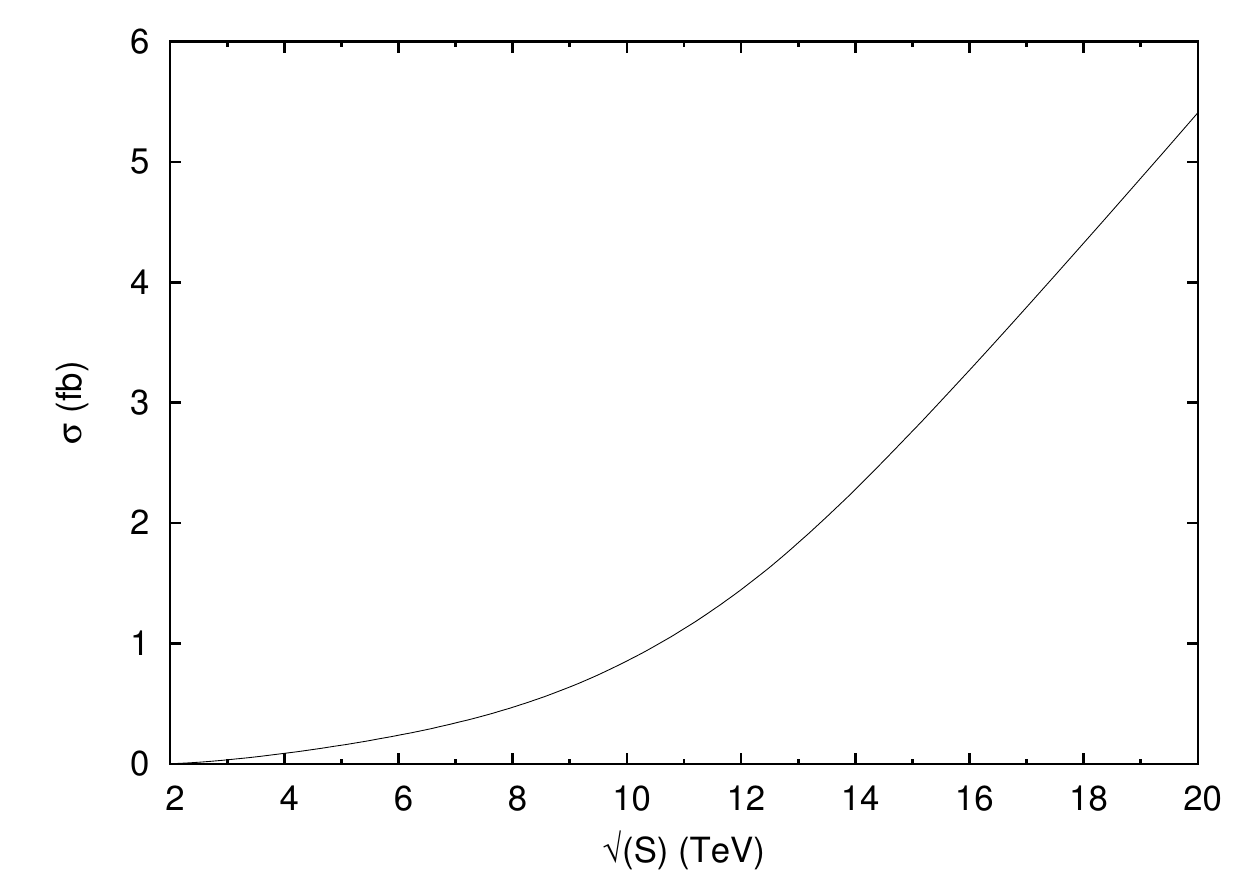}
\caption{Cross-section as a function of collider energy at the LHC.}
\end{center}
\end{figure}

\begin{figure}[!h]
\begin{center}
\includegraphics[width=8cm]{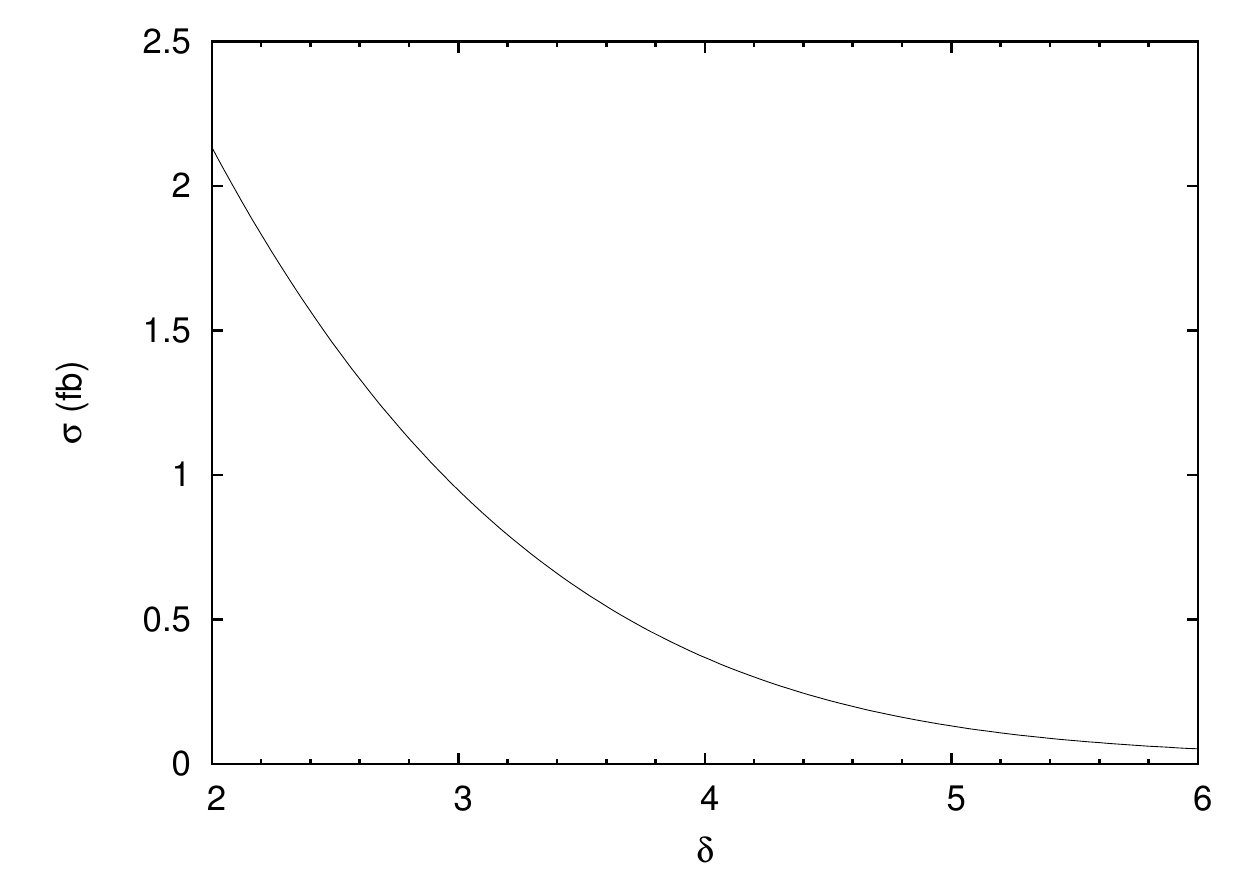}
\caption{Dependence of total cross-section on number of extra dimensions $\delta$, for $M_S = 2 $ TeV.}
\end{center}
\end{figure}

\begin{figure}[!h]
\begin{center}
\includegraphics[width=8cm]{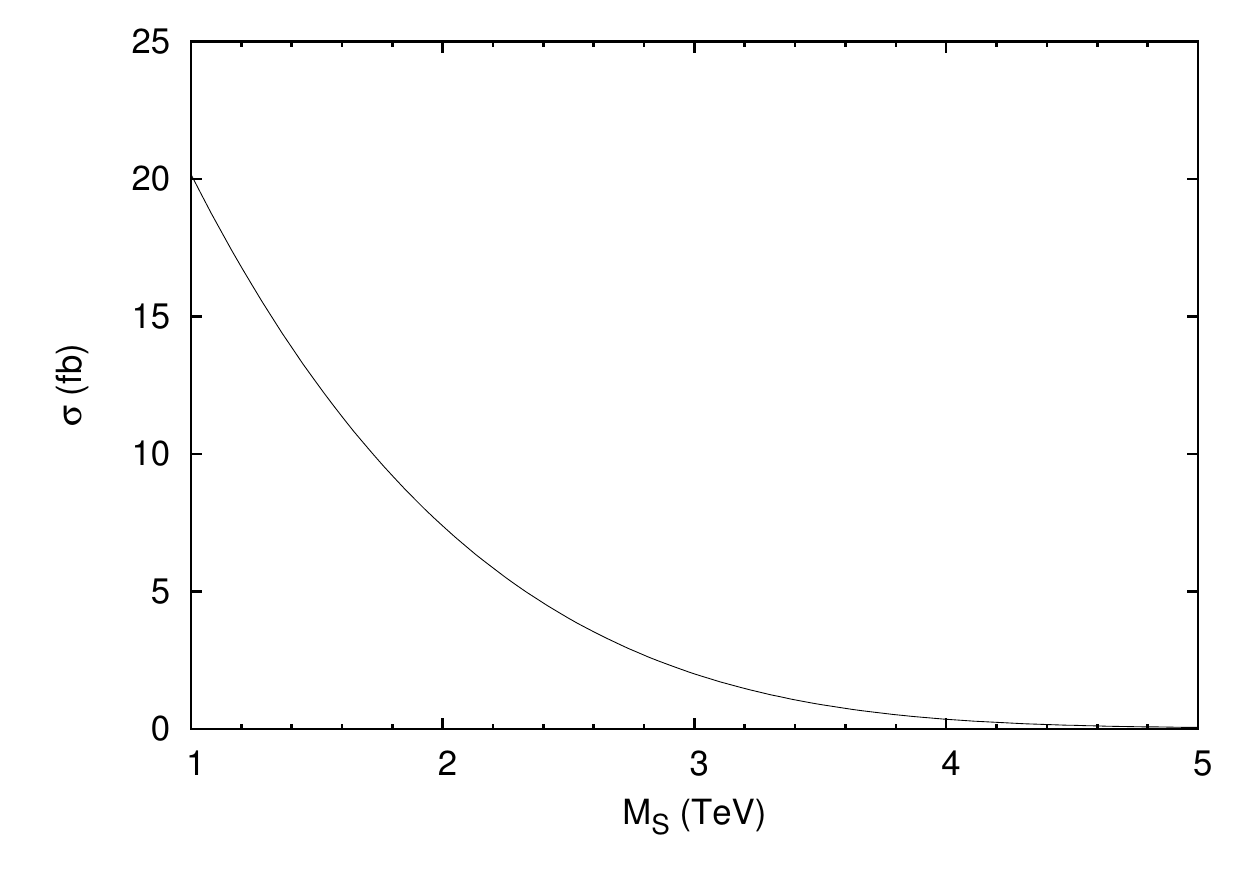}
\caption{Dependence of total cross-section on the scale  $M_S$, for $\delta = 2 $}
\end{center}
\end{figure}




\begin{figure}[!h]
\begin{center}
\includegraphics[width=8cm]{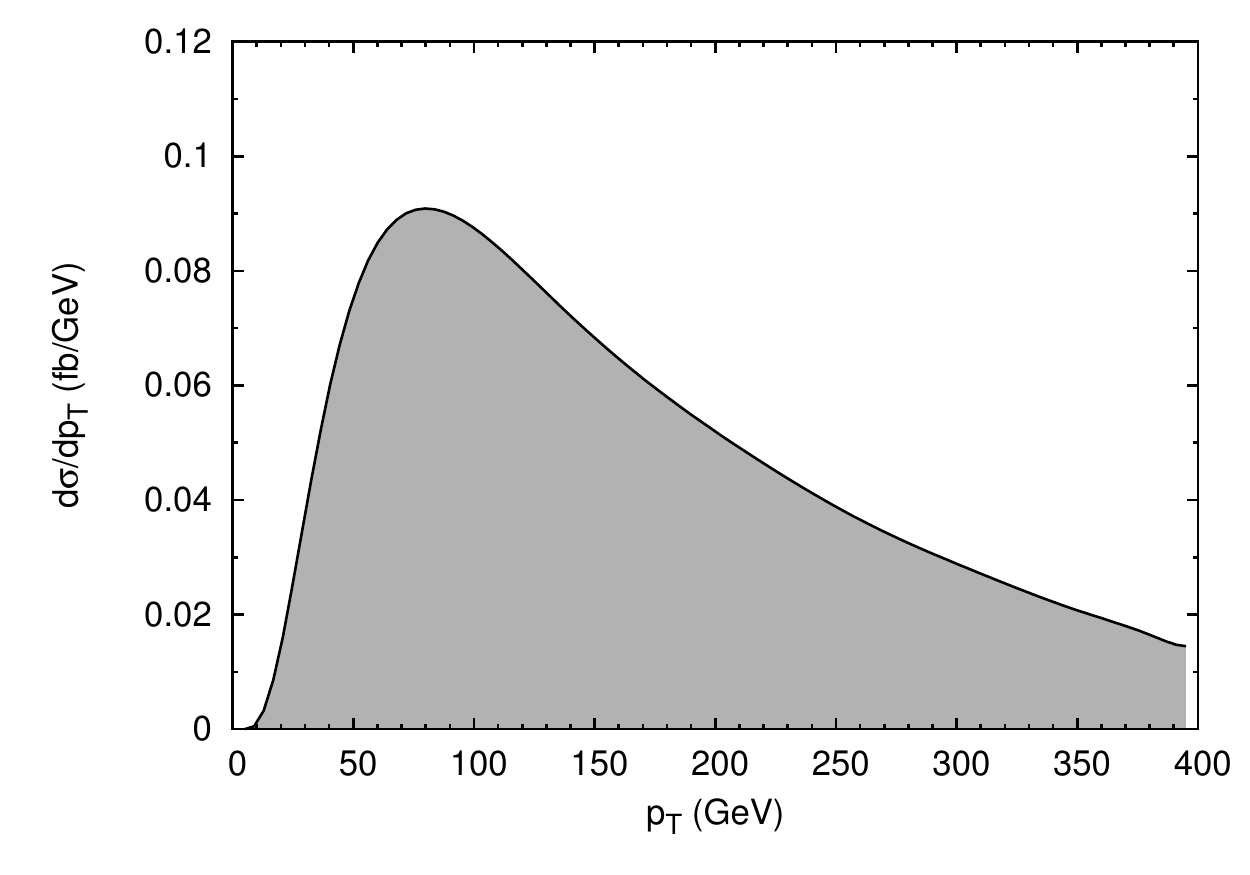}
\caption{Transverse momentum distribution of $Z$ for $M_S = 2$ TeV and $\delta = 2$.}
\end{center}
\end{figure}

\begin{figure}[!h]
\begin{center}
\includegraphics[width=8cm]{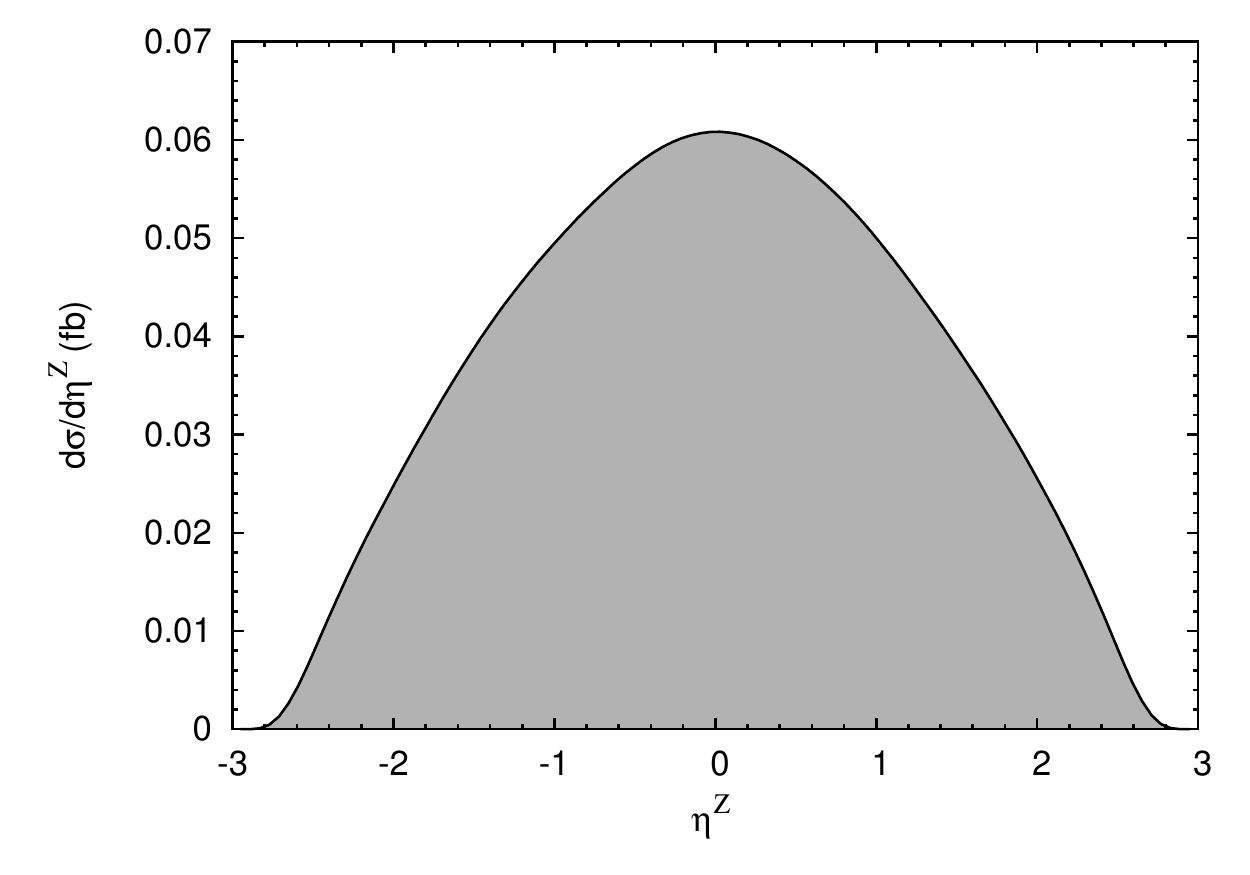}
\caption{Rapidity distribution of $Z$ for $M_S = 2$ TeV and $\delta = 2$.}
\end{center}
\end{figure}

\begin{figure}[!h]
\begin{center}
\includegraphics[width=8cm]{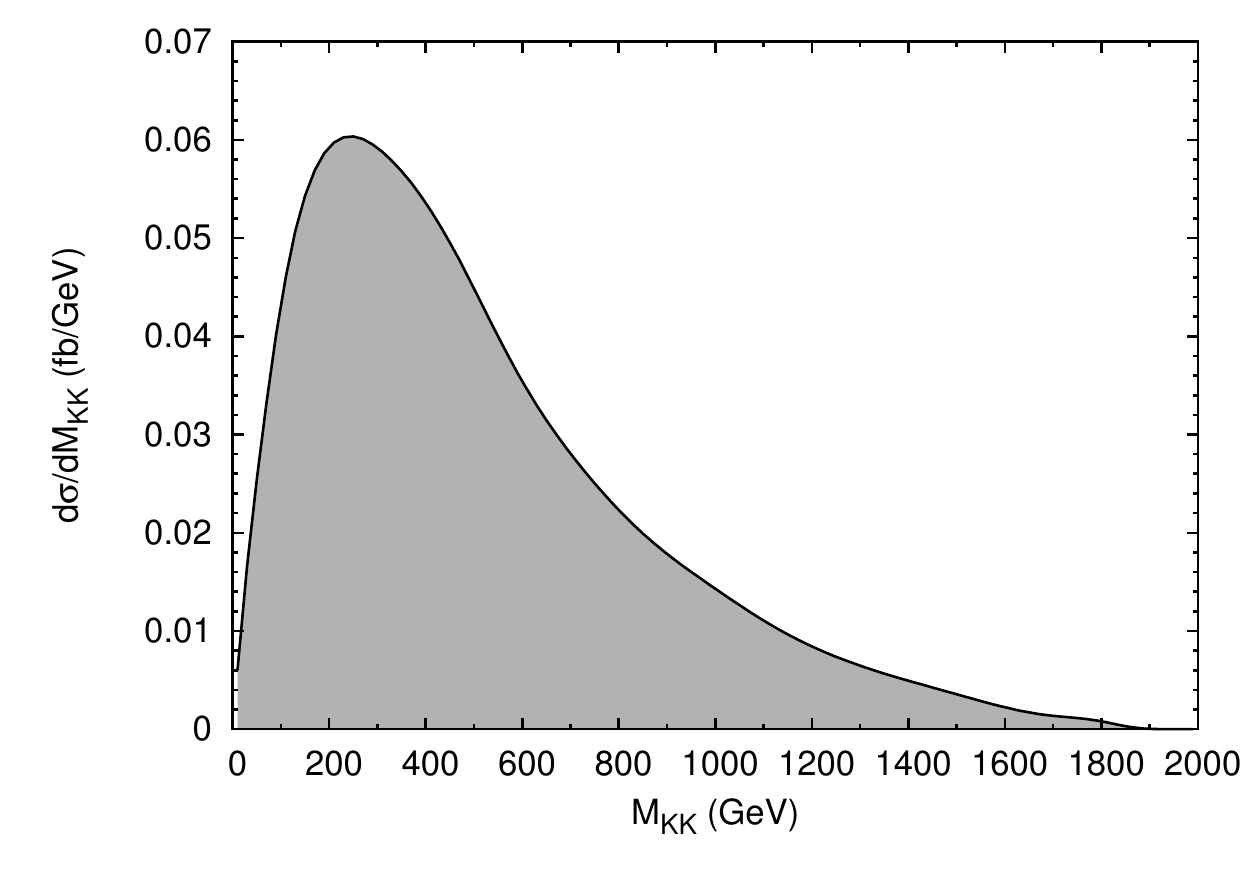}
\caption{KK-Graviton mass distribution for $M_S = 2$ TeV and $\delta = 2$.}
\end{center}
\end{figure}

In Figs. (3) and (4), we show the dependence of total cross-section on the two parameters of
 the ADD model, $M_S$ and $\delta$. As $\delta$ or $M_S$ is increased, the density of states for
 KK-modes falls and therefore the cross-section also goes down. 
The kinematic distributions of the transverse momentum and rapidity of $Z$ are plotted in 
Figs. (5) and (6), respectively.  Fig. (7) shows the KK-graviton mass distribution. Clearly,
the cross-section is peaked at a relatively larger value of about 200 GeV.
 We have also calculated the total cross-section for a $P_T^Z$ cut of 400 GeV
to avoid the SM background as suggested in \cite{Giudice:1998ck}. We find the cross-section 
of about $0.2$ fb and it's almost $10 \%$ of 
the NLO QCD correction calculated in \cite{Kumar:2010kv, Kumar:2010ca}. \\

Next, we study the scale dependence of the total cross-section. We vary the factorization scale 
around it's central value, $\mu_F = E_T^Z$. The cross-section
varies by about $30\% $ by changing $\mu_F$ in the range between $E_T^Z/2$ and $2E_T^Z$. 
The uncertainty in our calculation due to the choice of PDFs is between $5-20 \% $.
This is estimated by computing cross-section for other CTEQ parton distributions -- CTEQ6D,
CTEQ6L1, and CTEQ6L2. We have also studied the effect of taking large fermion mass limit 
in the loop. For a given phase-space point, we vary the top-quark mass and observe that 
the amplitude-squared (which includes both the bottom and top quark contributions) approaches 
a constant value as $m_t \rightarrow \infty$. This implies the complete decoupling of the
top quark, i.e.  the top quark contribution of the amplitude goes to zero in large $m_t$ limit.
It is expected from the decoupling theorem \cite{Appelquist:1974tg}. This feature has been plotted in Fig. (8).
The change in slope around $m_t = \frac{\sqrt{s}}{2} $ corresponds to physical threshold after which 
top-quark pairs can't be produced on-shell and the amplitude is real. \\

\begin{figure}[!h]
\begin{center}
\includegraphics[width=8cm]{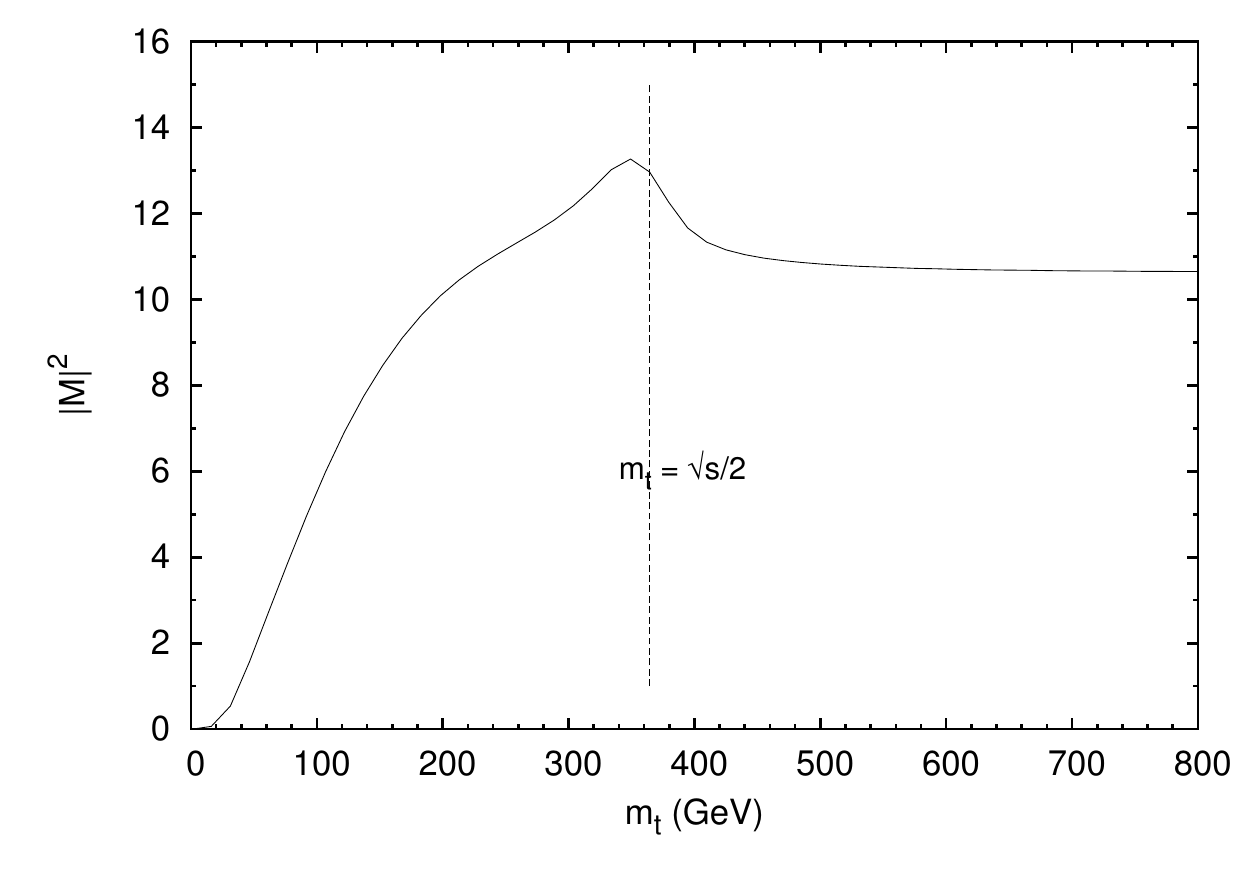}
\caption{Decoupling of $t$-quark in $m_t \rightarrow \infty$ limit.}
\end{center}
\end{figure}

Finally, we comment on the UV sensitivity of our theoretical results. As mentioned before, 
we have presented all the results for which the 
partonic energy $\sqrt{\hat{s}}$ is below the fundamental scale $M_S$, i.e. the truncated scheme. 
It has been argued in Ref. \cite{Giudice:1998ck} that, if relaxing 
this constraint does not change the results significantly, then results of the effective field theory
can be trusted. 
We find that the total cross-section 
in two cases (with and without cut on $\sqrt{\hat{s}}$) differs by $20 \%$ for $\delta=2$ and $M_S = 2$ TeV. 
This difference, as expected, increases 
for larger values of $\delta$, while decreases with increasing $M_S$.

\section{Conclusion}

We have presented the leading order calculation for the production of a KK-graviton in 
association with a vector boson via gluon fusion in the ADD model. Results have been presented 
for cross-sections and some important kinematic distributions at the LHC. 
We find that the amplitude for the $\gamma G$ case 
is zero at the LO, while in $Z G$ case only axial part of the amplitude contributes. We
have studied the qualitative features of our amplitude with respect to the anomaly 
structure and the decoupling of heavy fermions in the loop.
We saw that both the triangle and box diagrams give rise to anomaly
in the amplitude and this anomaly does cancel after considering contributions 
of other quark flavors. We also observed that the amplitude for a given quark flavor 
in the loop vanishes in the large quark mass limit verifying the decoupling theorem.   \\

The total cross-section for the gluonic contribution is small compared to the 
NLO results for the $ZG$ production at the LHC reported in \cite{Kumar:2010kv, Kumar:2010ca}. 
The smallness of the cross-section is due to the destructive interference between the
box-type and triangle-type diagrams. There is a relative negative sign between the
two contributions. This leads to the two-orders of magnitude cancellation that takes 
place between box and triangle contributions at the amplitude level.  Still, there can
be about 100 events per year when LHC will operate with peak CM energy and luminosity.

\section*{Acknowledgements}

AS would like to thank Prof. Lance Dixon, Prof. Tao Han, Prof. Geffory Bodwin, Prof. Frank Petriello,
 Prof. R. K. Ellis and Prof. J. Campbell for very fruitful discussions. He would also like to thank
the RECAPP center at HRI, Allahabad for supporting his visits.

\section{Appendix}

\subsection{Issue of $\gamma_5$ and Chiral Anomaly}
 
The  triangle amplitudes for our process have $VVA$-structure, 
while box amplitude has $VVVA$-structure. The issue of chiral anomaly 
is well known in linearly divergent fermion loop 
triangle diagrams with $VVA$-structure. Although box diagrams 
have $VVVA$-structure, they are also linearly divergent because of 
the nature of graviton coupling with quarks. Therefore, both the triangle 
and box diagrams will give rise to anomalous contribution.
The presence of anomaly affects the gauge invariance of the amplitude. 
For reliable predictions, our amplitude should be gauge invariant 
with respect to all the currents. Let's try to understand the issue 
of anomaly at the level of general structure of one loop amplitudes given in eqn. (8). \\

As discussed earlier, the rational term $\mathcal{R}$ in eqn. (8) is due to
the UV regularization of tensor integrals and it is 
independent of masses of internal lines. So, in dimensional regularization,
whenever $\frac{1}{\eps}$ UV-pole hits an $\mathcal{O}(\eps)$ term, we get a finite 
contribution. $\mathcal{R}$ is collection of all such terms obtained during 
the reduction of tensor integrals in $n=4-2\eps$ dimensions. It should be 
clear from the general structure of one-loop amplitudes, 
that in a gauge invariant amplitude $\mathcal{R}$ should be 
separately gauge invariant. In the case of one-loop calculation of 
fermion-loop diagrams in dimensional regularization,  the
issue of anomaly is related to the presence of $\gamma_5$ 
in a linearly (or worse) divergent integrals. There is always an 
ambiguity in carrying out full calculation in dimensional regularization,
as $\gamma_5$ is strictly a $4$-dimensional object. But this ambiguity 
arises only in fixing the rational part of the amplitude. Any $\mathcal{O}(\eps)$ 
structure of $\gamma_5$, if it at all exists in $n= 4-2\eps$ dimensions, will
contribute to the rational term as explained above. Thus anomalies affect 
only the rational part of the amplitude. They are also independent of the fermion mass. 
In the presence of anomalies, the gauge invariance of the amplitude is violated 
by these rational terms only. \\

The anomaly being associated with $\gamma_5$ should spoil only the axial-vector
current conservation and the conservation of vector current must hold.
But it is not very clear if this feature can always be ensured by treating $\gamma_5$ in $4$-dimensions. The $4$-dimensional $\gamma_5$ may 
lead to spurious anomalies in the amplitude, resulting in non-conservation of vector currents.
To ensure the vector current conservation in the amplitude,
it is advisable to use an appropriate prescription for $\gamma_5$ in $n$-dimensions. It is important to note that if we don't treat
$\gamma_5$ appropriately in $n$-dimensions, even the relation between charge conjugated fermion loop diagrams does not hold and as expected the 
violation is only in the rational terms. Thus by using an appropriate prescription for $\gamma_5$ in $n$-dimensions we generate a rational 
term consistent with various symmetries of the amplitude.
The $\gamma_5$ prescription given in eqn. (7) does this job. One should be careful using this prescription when generating other amplitudes 
by permuting external momenta and polarizations in prototype amplitudes. Any amplitude in which position of $\gamma_5$ should also change 
under such permutations should be computed separately. Thus within this prescription we require two prototype box and four prototype triangle 
(two for each class) amplitudes to generate the full amplitude. \\

The amplitude thus obtained for a given quark flavor in the loop is anomalous with respect to the $Z$-boson current only. As we said
earlier this anomaly in axial-vector current should also vanish. Since chiral anomaly is independent of quark mass and our amplitude is proportional
to the $T_3$ value of each quark flavor, the anomaly in axial-vector current does vanish when contributions from all six quark flavor is 
considered as in eqn. (6). \\ 

In the light of above understanding of anomalies, it is clear that even if 
we don't use any prescription for $\gamma_5$, as far as total amplitude in eqn. (6) is concerned, we must get a gauge
invariant result with respect to all the currents. Although amplitude for each quark flavor will not be gauge invariant any more,
their difference in eqn. (6) is always gauge invariant. This is because the rational terms being independent of 
quark mass cancel. We have 
verified this in a separate calculation and all the numerical results described in section 4 agree with this way 
of doing calculation. \\

\subsection{Decoupling of heavy fermions and Rational term}

In section 4, we found that if we use $n$-dimensional prescription for $\gamma_5$, the 
amplitude for a given quark flavor in the loop goes to zero as the large quark mass limit is taken. The decoupling of 
heavy fermions is, in general, expected in any UV finite fermion loop amplitude involving couplings 
which are not proportional to the fermion mass in the loop.
This is a very nice feature of fermion loop amplitudes. It can be utilized to fix the rational term of the amplitude 
without any ambiguity. It is particularly important in our calculation where the generation of correct rational term 
requires an $n$-dimensional prescription for $\gamma_5$ consistent with the vector current conservation. 
Since our amplitude is UV finite and decoupling of heavy fermions does hold, it implies that there must be a cancellation
between the mass independent rational term and $non$-rational part of the amplitude in eqn. (8), 
as large fermion mass limit is taken. Therefore, we do not need to calculate rational terms explicitly. 
The rational term $\mathcal{R}$ is simply the negative of the $non$-ration part of the amplitude 
in large fermion mass limit. Though decoupling does not hold if Higgs is attached to the fermion
loop, the structure of the fermion loop amplitude (with Higgs) suggests that one can
still fix the rational term by taking large fermion mass limit. For that one should take away the 
overall (explicit) factor of the fermion mass before taking the limit. Once again the amplitude
goes to zero indicating cancellation between rational term and $non$-rational part of the amplitude.

\end{document}